# Driver Behavior Post Cannabis Consumption – A Driving Simulator Study in Collaboration with Montgomery County, Maryland

Snehanshu Banerjee[a]*, Nashid K Khadem[b], Md Muhib Kabir[b] and Mansoureh Jeihani[b]

[a]*Omni Strategy LLC, Baltimore, Maryland, United States;*

[b]*Department of Transportation and Urban Infrastructure Studies, Morgan State University, Baltimore, Maryland, United States*

* Corresponding Author – sbanerjee@omni-strategy.com

# Driver Behavior Post Cannabis Consumption – A Driving Simulator Study in Collaboration with the Montgomery County

Montgomery County Police Department in Maryland hosted a cannabis intoxication impaired driving lab, to evaluate the driving behavior of medical marijuana users, pre and post cannabis consumption. A portable driving simulator with an eye-tracking device was used where ten participants drove a virtual network in two different events – a traffic light changing from green to yellow and the sudden appearance of a jaywalking pedestrian. An accelerated failure time (AFT) model was used to calculate the speed reduction times in both scenarios. The results showed that the participant speed reduction times are lower i.e., they brake harder post cannabis consumption, when they encounter a change in traffic light whereas there is no statistical difference in speed reduction times pre and post cannabis consumption, when they encounter a jaywalking pedestrian. The gaze analysis finds no significant difference in eye gaze pre and post cannabis consumption in both events.

1. **Introduction**

In contemporary times, drugged driving is a significant road safety concern worldwide. Several experimental, driving simulator and on-road driving studies have found that different drugs, including cannabis, debilitate driving performance, psychomotor abilities and cognition (Battistella et al., 2013; Lundqvist, 2005; Johannes G Ramaekers et al., 2004). Other studies show that 8.80 percent to 39.60 percent of road fatalities and 2.07 percent to 42.30 percent of road injuries happen due to drug-affected driving (Armstrong et al., 2018; Mura et al., 2006; Swann et al., 2004). Among drugs, cannabis or marijuana is frequently detected as an illicit drug in drivers' blood tests (Berning et al., 2015; Legrand et al., 2013; Pilkinton et al., 2013). One study shows that from 2007 to 2014, the amount of 9-tetrahydrocannabinol (THC)-positive drivers, increased by 48 percent on the weekends (Berning et al., 2015).

The strong relationship between drug use and crash likelihood becomes a prime concern for road safety (Asbridge et al., 2012; Drummer et al., 2004; Drummer et al., 2003; Mura et al., 2006). The crash risk associated with driving after acute cannabis use is not accurately or adequately presented. The data from the one crash risk study (Asbridge et al., 2012) has been re-analysed (Rogeberg & Elvik, 2016) yielding a lower estimate of risk, while other studies have had mixed results, including some particularly well-controlled studies with contradictory results (J. H. Lacey et al., 2016). It is clear from the literature that almost all the developed countries are encountering drugged driving problems. One study in Canada used multiple indicators collected from the Road Safety Monitor (RSM) and Canada's National Fatality Database maintained by the Traffic Injury Research Foundation (TIRF) to assess the trends of drugged driving (Robertson et al., 2017). The results revealed that the number of drivers driving after two hours of consuming cannabis increased from 1.6 percent in 2013 to 2.6 percent in 2015. The number of fatally injured drivers who tested positive for cannabis consumption also increased from 12.8 percent in 2000 to 19.7 percent in 2012 (Robertson et al., 2017). In New Zealand, cannabis is the most commonly detected illicit drug found in the blood tests of those who died in crashes (Poulsen et al., 2014; Poulsen et al., 2012). Drug usage other than cannabis is largely unknown in New Zealand (Poulsen et al., 2014; Poulsen et al., 2012). In Spain, a roadside survey shows that almost 17 percent of drivers test positive for alcohol and/or illicit drugs (Gómez-Talegón et al., 2012). Between 1991 and 2000, 52 percent of the blood samples of drivers who were involved in a fatal crash on the Spanish roads tested positive for alcohol and illicit drugs (del Río et al., 2002). A similar trend is found in the investigation of road traffic crashes across Europe. The drivers who tested positive for alcohol or illicit drugs range from 40 percent to 70 percent (Bogstrand et al., 2011;

Hamnett et al., 2017). In Switzerland, a study conducted by Senna et al. (Senna et al., 2010) found that 48 percent of suspected drug-impaired drivers tested positive for cannabis. Another meta-analysis, based on nine studies and 49,411 participants, concluded that under the influence of cannabis, the risk of a motor vehicle collision was almost twice that of regular drivers (Asbridge et al., 2012). However, car crash injuries appeared strongly associated with a strong cannabis intake rather than regular use (Blows et al., 2005). Illicit drugs are counterproductive to accomplishing a complicated task like driving and increase the probability of a crash (Elvik, 2013; Jongen et al., 2016; Johannes Gerardus Ramaekers et al., 2006; Ronen et al., 2010). In a study conducted in Milan, the concentration of cannabis and cocaine within legal limits had a correlation with car accidents more than legal blood alcohol concentration (BAC) (Ferrari et al., 2018). Some studies focus on the impact of age and gender on drug-related road accidents. Research shows that drugged driving occurs at various ages across both genders (Davey et al., 2014). Study results indicate that illicit drugs are more frequently found in young people whereas alcohol use is not age-related (Ferrari et al., 2018). Cannabis intake is common among young people whose median age is 27 while cocaine is frequent among adults in the median age group of 34 (Ferrari et al., 2018). However, research indicates that the highest prevalence of drugged driving is found among the age group of 18–30 (Akram & Forsyth, 2000), and males are represented more in injury statistics (Berry & Harrison, 2007; Drummer et al., 2004; Drummer et al., 2003; Longo et al., 2000).

In the United States, evidence suggests that the influence of cannabis is associated with a higher risk of road crashes (Ch'ng et al., 2007; Laumon et al., 2005; Lenné et al., 2010; Mann et al., 2010; Johannes Gerardus Ramaekers et al., 2006). In 2017, an estimated 122,943 Americans, 12 and older reported using cannabis (Statistics

& Quality, 2018). The usage was 34 percent in 2016, which is way higher than the 8.4 percent reported in 2014 (Sarra L. Hedden, 2015; Statistics & Quality, 2018). The increase in cannabis use could be a result of changes in state laws that allow medical or recreational use (Lipari et al., 2015). As of June 2019, 33 states and the District of Columbia permit the legal use of cannabis {Maciag, 2012 #302}, which can be seen in Figure 1.

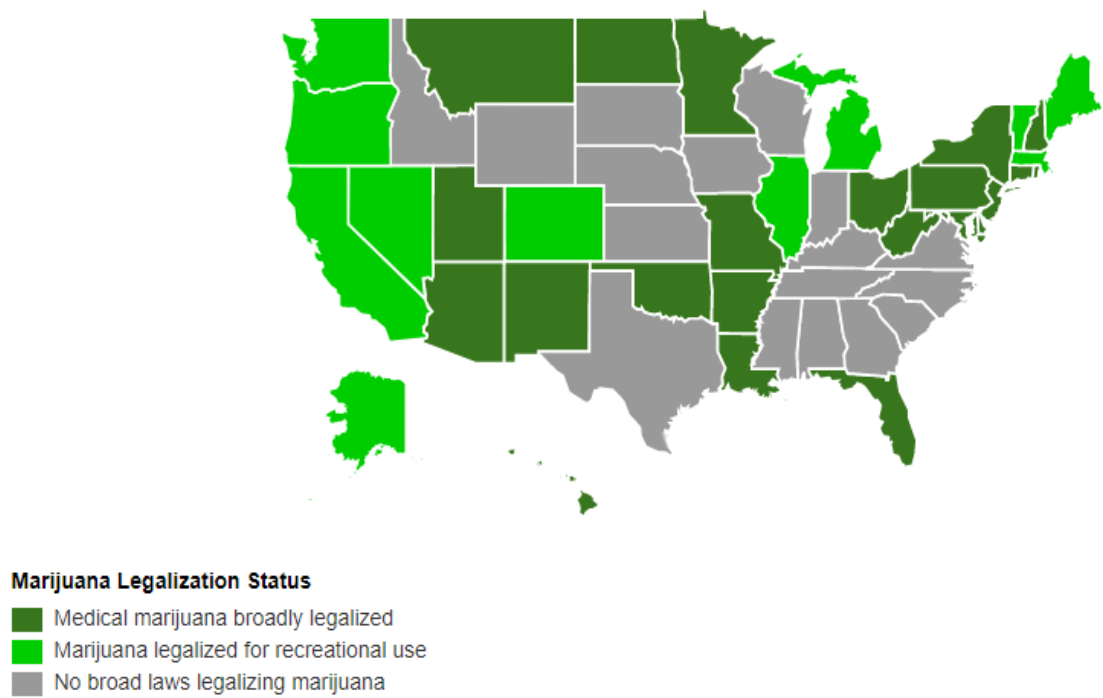

Figure 1. Legalization status of Cannabis – by State.

Smaller quantities of cannabis for personal consumption, which can range from one ounce to 10 grams, have been decriminalized in 21 states and Washington, D.C (M. Hartman, 2019; *State Medical Marijuana Laws*, 2019). These policy changes have positive and negative consequences on the health and safety of the general public. One important consequence of this is the impact of driving under the influence of cannabis. It is reported in past study that driving under the influence of drugs increases the chances of crashes and puts the passengers, driver and road users at risk (R. L. Hartman

& Huestis, 2013). Past research studies have shown consumption of cannabis impairs driving ability (R. L. Hartman et al., 2015; R. L. Hartman & Huestis, 2013), while others found moderate to no effect (Robbe, 1998). However, the pattern of road crashes has changed noticeably in some states after the legalization of cannabis. Colorado is experiencing more cases of driving under the influence of cannabis (DUIC) and fatal road crashes in comparison with the 34 states that haven't legalized medical cannabis (Salomonsen-Sautel et al., 2014). It is not easy to establish evidence-based law for DUIC as each state has different laws (Armentano, 2013; Grotenhermen et al., 2007; J. Lacey et al., 2010). Obviously, dosage of cannabis impact differs from person to person, as do intake methods and frequency of use (Azofeifa et al., 2015).

Driving simulator studies offer a chance to investigate the effect of cannabis on specific driving performance, and one study shows that a driving simulator is sensitive enough to reveal the THC-induced effects of higher dosages (Veldstra et al., 2015). Another driving simulator study showed that cannabis and alcohol increased the standard deviation of the lateral position (R. L. Hartman et al., 2015). Downey et al. (Downey et al., 2013) performed a driving simulator study in which driving performance was assessed at 20 min post-drug administration. Blood samples of the participants were taken before and after the performance. The results were obtained by ANOVA analysis and it was determined that participants who use cannabis regularly perform (drive) worse than infrequent cannabis users (Downey et al., 2013). Lenne et al. (Lenné et al., 2010) conducted a simulator study to understand the driving behavior under drug and alcohol influence. The driving performance was analyzed under nine different drug conditions in a driving simulator-designed arterial driving environment. The result shows that a high dosage of cannabis has a more significant impact on driving with regard to increased headway variability and decreased mean speeds, than a lower dosage. Ronen et al. (Ronen et al., 2010) performed another driving simulator study to understand the influence of alcohol and drugs on driving performance. Their results revealed that 3 out of 12 participants had a collision under THC. Smoking THC cigarettes causes a trend of slower driving among drivers and in combination with alcohol decreases subjects' ability to keep the steering wheel steady (Ronen et al., 2010). Liguori et al. (Liguori et al., 2002) used Post-hoc Tukey's pairwise comparison analysis to interpret significant interactions found in the simulator-based study. However, their study did not find any significant additive effects of alcohol and cannabis on brake latency or body sway (Liguori et al., 2002).

Recreational cannabis use is still illegal in Maryland. Medical marijuana has been legal in Maryland since 2014 (*State Medical Marijuana Laws*, 2019), but the authors haven't found any study in which medical marijuana users were evaluated for their driving behavior after consumption. The authors were unable to identify similar research done in the past and as such, this study is the first of its kind, evaluating the driving behavior of medical marijuana users post marijuana consumption at a police academy, using a portable driving simulator and an eye tracking device.

## 2. Methods

The Montgomery County Police Department in Maryland is the first police agency in the United States to host a cannabis intoxication impaired driving lab. This lab's primary purpose is to train police officers to better recognize cannabis impairment as it relates to driving. There are other for-profit organizations that conduct similar training; however, the authors were informed by the county police officials that there no other police agency currently conducts this type of training to better equip their patrol officers to make correct arrest decisions, when it comes to cannabis impaired driving. This study took place at the Montgomery County Public Safety Training Academy in Maryland under the supervision of Montgomery County police officers. The tools and methods used in this study have been explained in the following sections:

*2.1. Participants*

The Montgomery County Police Department collected the participant's data from different drug store and invite them for volunteer participation in a one-day pilot program. Ten adults (6 male and 4 females, ages 18 – 55, 50% white) are accepted the request to participate in driving simulator study. The participants had a history of consuming cannabis for medical conditions, ranging from a year up to 10 years. Some 80% of the participants answered that they consume cannabis several times a day, while the remaining 20% said they consume cannabis only 6-7 times a week.

Table 1. Cannabis related descriptives.

|  | Quantity | | THC | | Ingestion Method | | Oral Fluid | |
| --- | --- | --- | --- | --- | --- | --- | --- | --- |
|  | 1st Dose | 2nd Dose | THC (1) | THC (2) | Smoked | Vape | Pos. | Neg. |
| **Participant 1** | 5 dabs | 1.5 gms | 88.5 | 26.95 | X (2) | X (1) | X |  |
| **Participant 2** | 0.5 gms | 0.25 gms | 25.1 | 83.82 | X (1) | X (2) | X |  |
| **Participant 3** | 0.1 gms | 0.05 gms | 88.05 | 88.05 |  | X (1 & 2) | X |  |
| **Participant 4** | 1.5 gms | 0.8 gms | 23.7 | 23.7 | X (1&2) |  |  | X |
| **Participant 5** | 3 pulls | 2 pulls | 77.3 | 77.3 |  | X (1 & 2) |  | X |
| **Participant 6** | 1.6 gms | 1 gm | 33.4 | 33.4 | X (1&2) |  |  | X |
| **Participant 7** | 0.2 gms | 0.3 gms | 76.3 | 76.3 |  | X (1 & 2) | X |  |

| | | | | | | |
|---|---|---|---|---|---|---|
| **Participant 8** | 5 pulls | 4 pulls | 77.3 | 0/CBD only | X (1 & 2) | X |
| **Participant 9** | 0.5 gms | 0.5 gms | 28.09 | 28.09 | X (1 & 2) | X |
| **Participant 10** | 0.5 gms | 0.5 gms | 76.7 | 76.7 | X (1 & 2) | X |

CBD – cannabidiol, similar to THC but has a non-psychoactive compound.

dabs – concentrated dose of cannabis made by extracting THC/other cannabinoids using butane or $CO_2$ resulting in sticky oils.

All of the participants said that cannabis helps them relax while 90% of the participants mentioned that it can help people with mental illnesses. Tetrahydrocannabinol (THC) is the main psychoactive ingredient in cannabis. To make the study more realistic, researchers used cannabis with THC similar to what is available on the streets. The percentage of THC present in the plant adds to the impairment. Hypothetically a lower THC strain should impair a person less. After driving the baseline scenario – i.e., pre cannabis consumption –participants were dosed twice within a period of two hours. The columns dictating THC (1) or THC (2) in Table 1 list the percentage of THC present in the first dose (1) or the second dose (2). The participants were given an oral fluid test prior to the driving start of the study. They were instructed not to consume cannabis that day; however, based on their usage (or inability to follow the rules) they tested positive on the oral fluid test. This only means that within their oral fluid they met the threshold of have 25 nanograms or higher in their saliva. Although this is very different than nanograms in the blood and is not significant, it is noteworthy. The dosage, the potency, method of ingestion and test results are shown in Table 1 and were tested for correlations later in the study.

*2.2. Driving Simulator and Network Setup*

This study used a portable driving simulator, which includes a TMX steering wheel with a 900-degree force feedback base and a brake and gas pedal (Figure 2), connected to a 65-inch TV screen. This equipment was used rather than the two medium fidelity driving simulators owned by the Safety and Behavioral Analysis Center (SABA) at Morgan State University, as this study had to be conducted off campus at the Montgomery County Public Safety Training Academy. The portable simulator was backed by the software VR-Design Studio developed by FORUM8 Co (*Advanced Driving Simulation*). Driving behavioral data such as braking, acceleration, steering wheel control and speed were recorded in real time using this software.

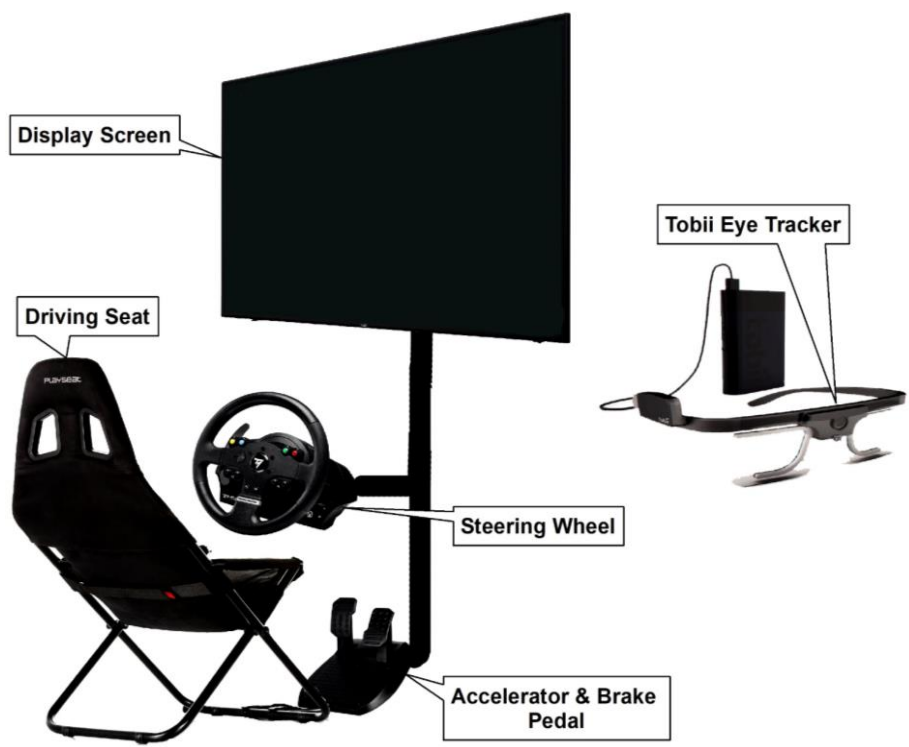

Figure 2. Study apparatus

A network was built replicating an area south of Baltimore that connects to downtown Baltimore: a two-lane 1.1-mile highway, Interstate 395 with a speed limit of 40 mph that reduces to 30 mph as it approaches downtown Baltimore, in the Inner Harbor area. Two events were designed in the same scenario, which were tested in both the pre and post cannabis consumption phases, which occurred with a gap of a minimum of two hours.

*2.2.1. Change in Traffic Light*

The two-lane highway, interstate 395, has a speed limit of 30 mph as it approaches the Conway Street signalized intersection, near downtown Baltimore. The lanes are 12 feet wide with a 12-foot raised median separating the opposing lanes with a light traffic flow with a level of service B. In this event, when the participant reaches an arbitrary distance of 50 meters from the traffic signal stop line, the traffic light changes from green to yellow. The distance of 50 meters was chosen based on topography, road geometry and feedback from five random people, who were invited to give their perspectives on object visibility, i.e. when does the object, the traffic light in this case, become visible to them. Although the authors attempted to follow the Manual on Uniform Traffic Control Devices (MUTCD), to establish minimum distances for objects based on the guidelines, since this is a driving simulator, the visibility differs from real life. A snapshot of this event is shown in Figure 3.

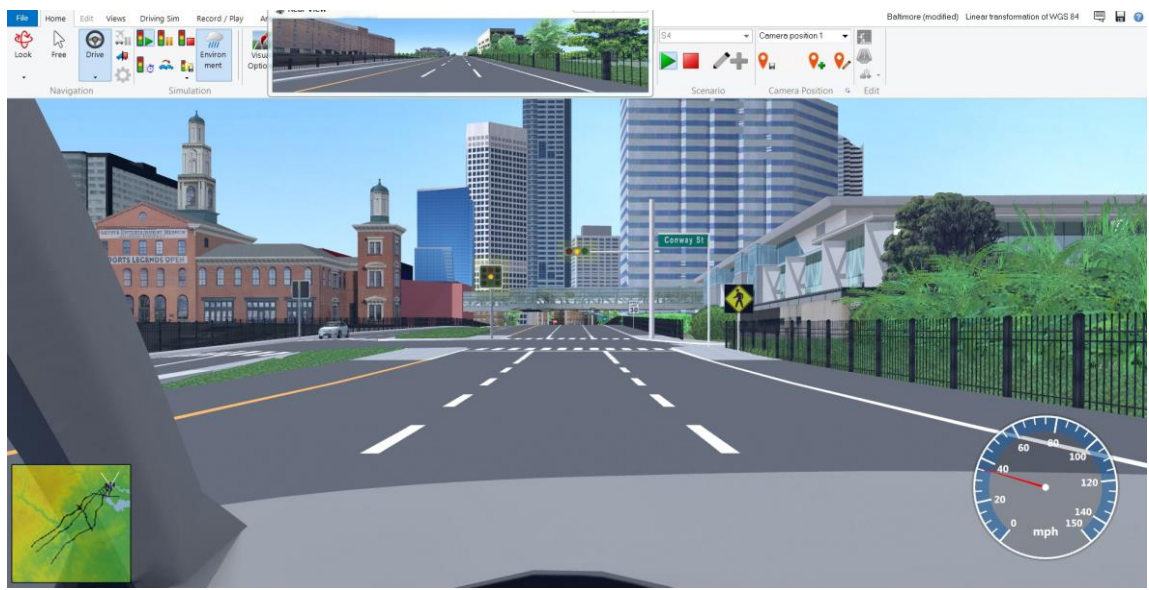

Figure 3. Change in traffic light.

*2.2.2 Sudden Appearance of a Jaywalking Pedestrian*

Pratt Street, a four-lane major one-way road in downtown Baltimore with a speed limit of 30 mph, was used for this event, as it has a lot of foot traffic. Traffic flow was light, with a level of service B. It is a complete street, with a 14-foot-wide wide shared bus and bike lane, three 12-foot lanes and wider sidewalks. For this event, as the participant drives east, a pedestrian starts jaywalking approximately 40 meters in front of the participant. The distance of 40 meters was chosen based on traffic conditions, road geometry and feedback from five random people, who were invited to give their perspectives on object visibility, i.e., when does the object, pedestrian in this case, become visible to them. Although the authors attempted to follow the Manual on Uniform Traffic Control Devices (MUTCD), to establish minimum distances for objects based on the guidelines, since this is a driving simulator, the visibility differs from real life. A snapshot for this event is shown in Figure 4.

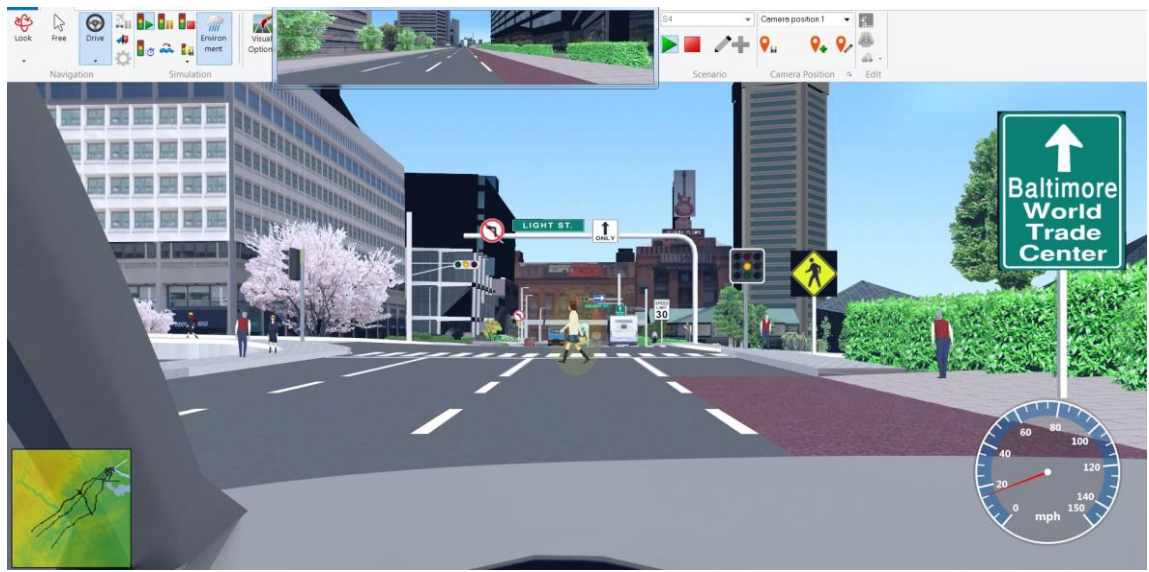

Figure 4. Jaywalking pedestrian.

Two pre and post scenario survey questionnaires were developed to gather information on sociodemographics, cannabis consumption-related questions and simulator driving experience questions. Montgomery County police had already obtained approval to test human subjects for this study.

## 2.3. Hazard-based Duration Model

Hazard-based duration models have been used to study a number of time-related events such as driver reaction time when using a cell phone while a pedestrian crosses the street (Haque & Washington, 2014, 2015), probing delays in traffic due to frequent incidents, blocked lanes (Qi et al., 2009), duration modeling for highway traffic incidents (Hojati et al., 2013; Junhua et al., 2013; Nam & Mannering, 2000), etc.

The duration variable in this study is the speed reduction time, which is calculated at the onset of an event, until the minimum speed is reached before the participant starts accelerating again. The conditional failure rate is given by the hazard function $h(t)$. Two parametric approaches could have been used in this study, a proportional hazard and an accelerated failure time (AFT) model. These models enable the researchers to analyze the impact of covariates on the hazard function. Compared to the hazard model, the AFT model enables the covariates to accelerate time in a survivor function when all covariates are zero as well as making the results easier to interpret, while the hazard ratios are constant over time {Simon P. Washington, 2011 #301}. Thus, an AFT modeling approach was chosen for this study.

## 3. Results and Discussion

Researchers built a parametric survival model to analyze an aspect of driver behavior, average deceleration rate in response to a traffic light changing from green to yellow and used an ANOVA analysis to compare speed reduction times in response to a jaywalking pedestrian appearing suddenly in front of the user vehicle.

*3.1. Change in Traffic Light*

The participants' braking behavior have been analyzed at the onset of the yellow light while attempting to stop. The speed profiles of each participant while approaching the stop line were analyzed, and variables related to the braking behavior or average deceleration rate have been plotted as shown in Figure 3. Speed reduction times were calculated from the moment the traffic signal light turned from green to yellow, i.e. the initial speed to the minimum speed reached before or at the stop line. Only 15 of the 20 observations were used to compute speed reduction times, as the others did not stop at the red light. It can be seen that the participants braked harder post cannabis consumption, i.e. they were more aware of the change in traffic light.

The average deceleration rate ($d_m$) at the onset of the yellow light until they attain minimum speed before the stop line is given by (Bella & Silvestri, 2016);

$$d_m = \frac{V_i^2 - V_{min}^2}{2(L_{V_{min}} - L_{V_i})}$$

where,

$V_i$ = Participant's initial speed as they approach the signal upto the point when the signal turns yellow

$V_{min}$ = Participant's minimum speed reached during the deceleration phase before acceleration

$L_{V_i}$ = Distance on the road at which the signal turns yellow

$L_{V_{min}}$ = Distance from the stop line where the driving speed is minimum before acceleration

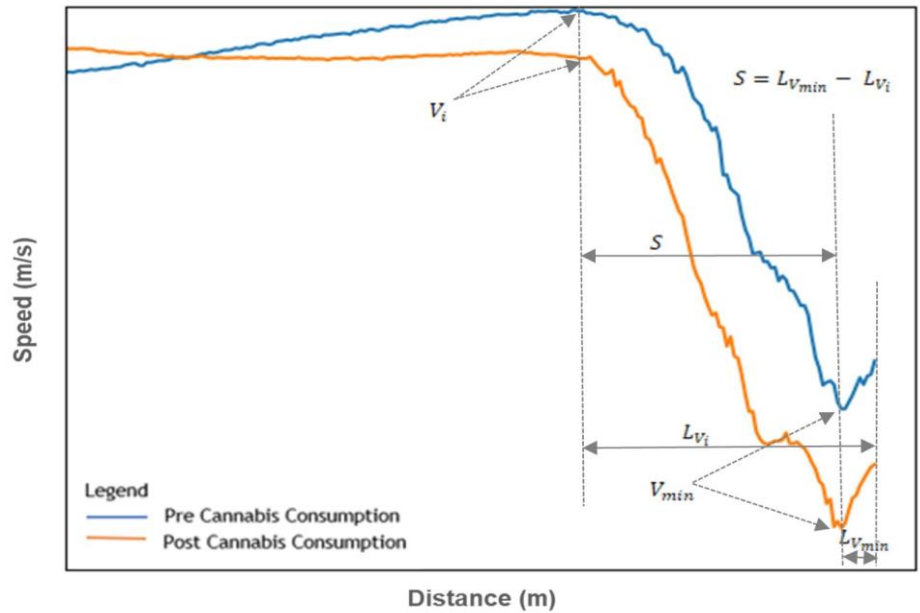

Figure 5. Participant braking behavior comparison.

The speed reduction time (S) is computed as the time elapsed between the participant's initial speed ($V_i$) and the minimum speed ($V_{min}$) reached before accelerating.

An ANOVA analysis revealed that there was a statistically significant difference in speed reduction times pre and post cannabis consumption. The speed reduction times were shorter (hard-braking) post cannabis consumption (mean difference in time = 0.82 seconds, $\rho \leq 0.05$) at 3.84 seconds compared to 4.66 seconds pre cannabis consumption. Since deceleration rate or braking behavior is affected by cannabis consumption, a hazard-based duration model was developed to comprehend the participant's braking behavior.

An AFT model analysis was appropriate for this study as the AFT models can assume that the effect of cannabis consumption could adversely affect life if a braking event doesn't occur, possibly resulting in a crash. Four different AFT distributions were tested, namely, the Weibull, log-logistic, log-normal and exponential to analyze the braking behavior. The maximized likelihood and Akaike information criterion (AIC) values were used to select the best model for this dataset. This study found the Weibull AFT distribution to be the best fit, since it has the lowest AIC and the highest maximum likelihood, for this dataset to evaluate braking behavior at the onset of the yellow light.

For the Weibull AFT model, the hazard functions h(t) and the survival functions S(t) are given by (Bella & Silvestri, 2016) in Equation 1 and 2:

$$h(t) = (\lambda P)(\lambda t)^{P-1} \tag{1}$$

$$S(t) = exp(-\lambda t)^P \tag{2}$$

where,

$\lambda$ = location parameter

P = Scale parameter

t = Specified time

After introducing explanatory variables, the location parameter can be expressed as shown in Equation 3:

$$\lambda = \exp\left[-P(\beta_0 + \beta_1 X_1 + \cdots)\right] \tag{3}$$

where the explanatory variables $X_i$ have the coefficients $\beta_i$.

The descriptive statistics of the Weibull AFT model are shown in Table 2.

Table 2. Speed reduction time and Weibull AFT variable descriptives

| Variables | Mean Value (pre cannabis consumption) | Std. Dev (pre cannabis consumption) | Mean Value (post cannabis consumption) | Std. Dev (post cannabis consumption) |
|---|---|---|---|---|
| $V_i$ (m/s) | 15.05 | 2.31 | 14.28 | 3.50 |
| $L_{V_i}$ (m) | 50.65 | 0.63 | 50.72 | 0.44 |
| $V_{min}$ (m/s) | 2.33 | 0.00 | 1.41 | 3.17 |
| $L_{V_{min}}$ (m) | 103.54 | 7.86 | 92.22 | 9.14 |
| $d_m$ (m/s²) | 2.02 | 0.85 | 2.43 | 0.82 |
| Speed Reduction Time(s) | 4.66 | 0.59 | 3.84 | 0.71 |

The Weibull AFT model can be assessed for its goodness of fit and demonstrated using a Cox-snell residual plot, which was determined from the model estimates and then used to build an empirical estimate of the cumulative hazard model. Although the data points are limited, the predicted speed reduction time of the participants using this model matches well with the observed data as shown in Figure 6.

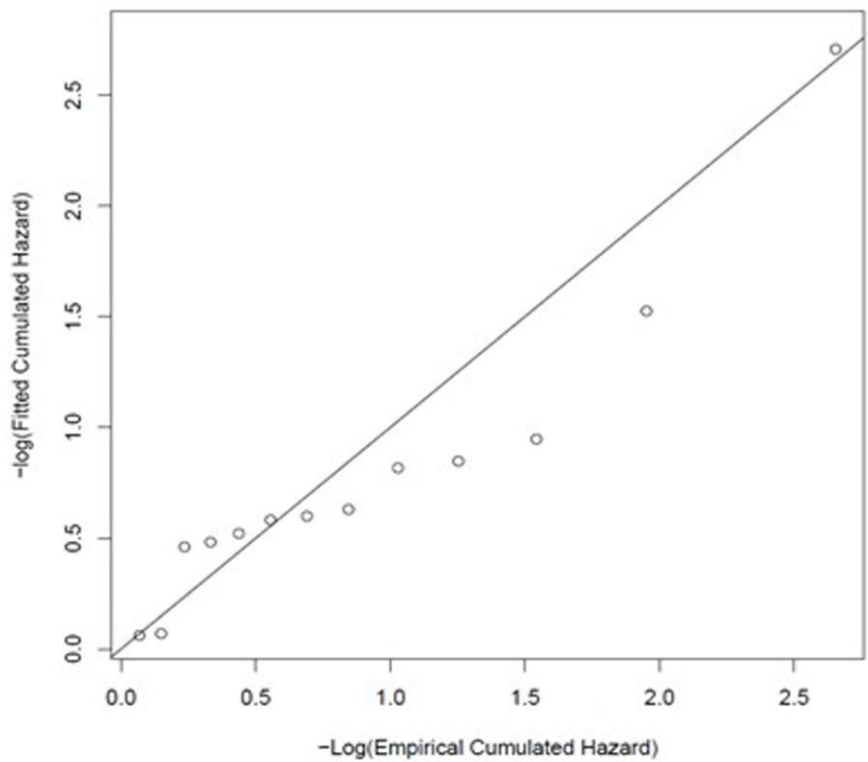

Figure 6. Cox-snell residuals for Weibull AFT.

The estimates from the Weibull AFT model with the speed reduction times of the participants as the dependent variable is shown in Table 3.

Table 3. Weibull AFT parameter estimates

| Variables | Estimate | Std. Error | z | ρ |
|---|---|---|---|---|
| (Intercept) | 38.979 | 131.988 | 0.295 | 0.768 |

| Variables | Estimate | Std. Error | z | ρ |
|---|---|---|---|---|
| $V_i$ (m/s) | 0.063 | 0.027 | 2.300 | 0.021* |
| $L_{V_i}$ (m) | -0.018 | 0.064 | -0.286 | 0.775 |
| Average Deceleration Rate $d_m$ (m/s²) | -0.333 | 0.086 | -3.854 | 0.000* |
| Pre cannabis consumption | 0.113 | 0.047 | 2.404 | 0.016* |
| Consumption increases heart rate | -0.134 | 0.152 | -0.881 | 0.378 |
| Consumption causes addiction | 0.013 | 0.139 | 0.094 | 0.925 |
| Consumption leads to reduced attention span | 0.070 | 0.104 | 0.672 | 0.501 |
| Participants driving immediately post cannabis consumption | 0.058 | 0.043 | 1.360 | 0.174 |
| Scale Parameter P | 3.806 | 0.244 | | |
| Log-likelihood at convergence | -14.541 | | | |
| AIC | -41.082 | | | |
| Number of observations | 15 | | | |

* Statistically significant at 95% CI

The model identifies the initial driving speed, the average deceleration rate and the pre cannabis consumption scenarios as significant factors (ρ ≤ 0.05) impacting speed reduction times. The coefficient of the average deceleration rate is negative, which indicates that the lower the average deceleration rate, the higher the speed reduction time. A scale parameter estimate of 3.806 implies that the survival/crash probability of the speed reduction times decreases with the passage of time. This means that the probability of the participant's response to the changing of the traffic light to yellow increases, i.e., the participant engages in a smoother braking maneuver over elapsed time. However, this condition is true only for the pre cannabis consumption scenario, in which the speed reduction time was lower by 0.82 seconds as seen in Table 2. This means that post cannabis consumption, the participants engage in aggressive braking manoeuvres resulting in decreased speed reduction times compared to pre cannabis consumption.

A representation of the participants' braking patterns can be shown by plotting survival curves of the speed reduction times for the pre and post cannabis consumption scenarios. These predictions were done based on the predict survival regression tool in the R-package (TM, 2019).

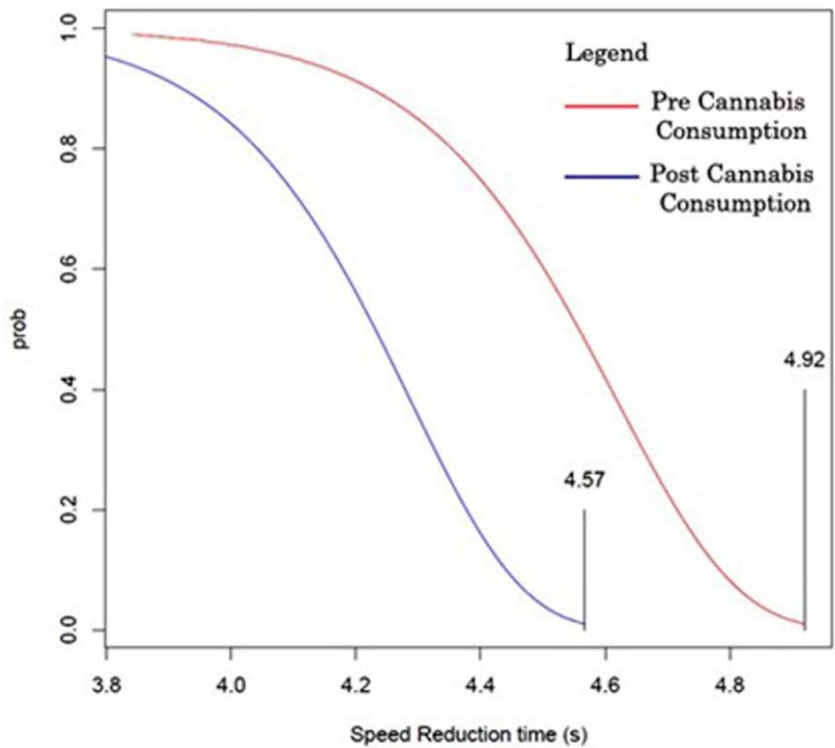

Figure 7. Speed reduction time survival curves.

From Figure 7, the speed reduction time survival/crash probability for the yellow light dilemma, decreases with the passage of time. A lower survival/crash probability was recorded for the post cannabis consumption scenario as compared to the pre cannabis consumption scenario. After 4 seconds of speed reduction time, the survival probability for post cannabis consumption was 84% compared to 97% for pre cannabis consumption and it drops even further at 4.5 seconds, to only 4% post cannabis consumption to 60% pre cannabis consumption. The speed reduction time was 0.35 seconds longer (statistically significant) for the pre cannabis consumption scenario compared to post cannabis consumption, thus giving the participants longer time to brake and come to a smooth stop. No correlation was found between sociodemographics of the participants and speed reduction times. An interesting observation was that 60% (6/10) of the participants stopped for the red light before as well as after cannabis consumption, while 30% (3/10) of the participants stopped for the red light only after cannabis consumption. Approximately 67% (2/3) of participants had a high THC concentration (> 75% on the 2nd dose), which could possibly indicate that a higher dosage of cannabis keeps certain medical marijuana users more alert.

### *3.2. Sudden Appearance of Jaywalking Pedestrian*

The observations where the participants were either speeding and missed the jaywalker or the pedestrian was run over, were not used in the analysis. The researchers attempted to build another parametric survival model for this event, but an ANOVA analysis conducted to analyze the speed reduction times once the jaywalking pedestrian appeared revealed no statistically significant difference in speeds in the pre and post cannabis consumption scenarios. Since no statistical significance in speed reduction times were found, the outputs were not included for this analysis. The speed reduction times were 3.2 seconds and 3 seconds for the pre and post cannabis consumption scenarios. An interesting observation was that 50% (5/10) of the participants stopped for the pedestrian only in the post cannabis consumption scenario and 60% (3/5) of these participants had a high THC concentration, which affirms what the authors found in the change in traffic light event, that a high dosage of THC, keeps certain medical marijuana users more alert.

### *3.3. Eye Gaze Analysis*

The eye tracking device Tobii Pro Glasses 2 (*Tobii Pro Glasses 2*) and its analysis software was used for this analysis. Both the events, change in traffic light and a jaywalking pedestrian, were analyzed for distractions. If the participant glances at an object for a moment, it is captured in the eye tracking device, which in this study was either the traffic light or the pedestrian in both pre and post cannabis consumption scenarios. An eye gaze heat map analysis is shown in Figure 8, which has been scaled for better visualization due to the low number of samples. The eye gaze at the change in traffic light event during the pre and post cannabis consumption is slightly different. This could possibly be attributed to the braking behavior post cannabis consumption, where participants braked harder compared to the pre cannabis consumption scenario. The gaze does not change significantly during the jaywalking pedestrian event in the pre as well as post cannabis consumption scenarios, with almost an equal number of participants being distracted pre and post cannabis consumption.

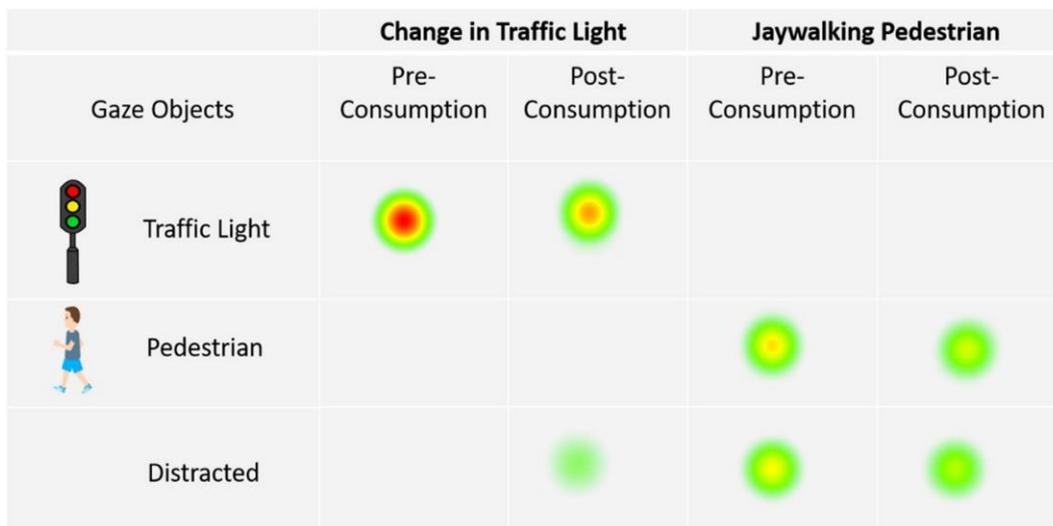

Figure 8. Eye tracking gaze analysis.

During the traffic light event, although almost all the participants noticed the traffic light changing, the authors observed that 60% of the participants who stopped at the red light, stopped far away i.e., around 8 – 10 meters from the stop line. This could possibly be attributed to the sudden braking behavior outcome from the Weibull AFT model, in the post cannabis consumption scenario. During the jaywalking event, it is observed that although there were distracted participants during the pre-cannabis consumption scenario, there was almost an equal number of participants who were distracted post cannabis consumption. The authors did not find any significant correlation with sociodemographics of the participants and eye gaze analysis.

**4. Conclusion**

This study investigated the influence of cannabis on driver behavior using a portable driving simulator. A total of 10 simulation sessions were conducted for 10 participants pre cannabis consumption and another 10 sessions post cannabis consumption, which consisted of two doses within a period of two hours. Two events were evaluated for speed reductions times, a changing traffic light and sudden appearance of a jaywalking pedestrian. It was found that speed reduction times were lower post cannabis consumption in the event of the traffic light changing from green to yellow. This possibly means that participants brake aggressively before coming to a stop at the red light. This transition is not smooth as observed in the eye tracking videos, in which 60% of the participants reduced their speeds way before the traffic signal, before inching to the stop line and coming to a final stop. This agrees with prior studies, in which consumption of THC resulted in slower speeds among drivers (Lenné et al., 2010; Ronen et al., 2010). The sudden appearance of a jaywalking pedestrian did not appear to have any statistically significant change in speed reduction behavior of the participants. This could possibly mean that, as medical marijuana users, they are still alert post consumption. Although the sample size for this study was insufficient to draw concrete conclusions, the authors observe that consumption of cannabis seems to cause no significant driving impairment to the participants in terms of speed reduction time, in the jaywalking pedestrian scenario. This agrees with a prior study in which no significant effects were found in braking behavior post cannabis consumption (Liguori et al., 2002). However, this study does not take into account cannabis consumption for everyday users without a history of cannabis consumption and its impact on their driving behavior, post cannabis consumption. A limitation of this study is the small sample size of participants and further research is needed with a higher sample size of participants; caution should be exercised, if the users decide to drive post cannabis consumption. Additionally, research must be carried out on people who use marijuana recreationally, as recreational use of marijuana is still illegal in many states including Maryland. As such, for recreational users, consuming cannabis and driving would possibly impact them differently than it does with medical marijuana users. Further research is needed to support this hypothesis and will be a focus for future studies.

Acknowledgements, the authors would like to thank Montgomery County police officials for organizing the study and inviting the authors to be an integral part of it. The authors would also like to thank the Maryland State Highway Administration, the Federal Highway Administration (FHWA) and the Governors Highway Safety Association (GHSA) officials who were present during the study.